\documentclass[final,5p,times,twocolumn]{elsarticle}

\usepackage{graphicx}
\usepackage{amssymb}
\usepackage{amsmath}
\usepackage{xcolor}

\begin{document}
	
	\begin{frontmatter}
		
		\title{Learning Hidden Chemistry with Deep Neural Networks}
		
		
		\author[a]{Tien-Cuong Nguyen}
		\author[b]{Van-Quyen Nguyen}
		\author[c]{Van-Linh Ngo}
		\author[d]{Quang-Khoat Than}
		
		\author[b,d]{Tien-Lam Pham\corref{mycorrespondingauthor}}
		\cortext[mycorrespondingauthor]{Corresponding author}
		\ead{lam.phamtien@phenikaa-uni.edu.vn}
		
		\address[a]{VNU-University of Science, 334 Nguyen Trai, Thanh Xuan Dist., Hanoi, Vietnam}
		\address[b]{Phenikaa Institute for Advanced Study (PIAS), Phenikaa University, Yen Nghia, Ha Dong Dist., Hanoi 12116, Vietnam}
		\address[c]{Hanoi University of Science and Technology, 1 Dai Co Viet, Hai Ba Trung Dist., Hanoi, Vietnam}
		\address[d]{Faculty of Computer Science, Phenikaa University, Yen Nghia, Ha Dong Dist., Hanoi 12116, Vietnam}

		\begin{abstract}
			We demonstrate a machine learning approach designed to extract hidden chemistry/physics to facilitate new materials discovery. In particular, we propose a novel method for learning latent knowledge from material structure data in which machine learning models are developed to present the possibility that an atom can be paired with a chemical environment in an observed materials. For this purpose, we trained deep neural networks acquiring information from the atom of interest and its environment to estimate the possibility. The models were then used to establish recommendation systems, which can suggest a list of atoms for an environment within a structure. The center atom of that environment was then replaced with the various recommended atoms to generate new structures. Based on these recommendations, we also propose a method of dissimilarity measurement between the atoms and, through hierarchical cluster analysis and visualization using the multidimensional scaling algorithm, illustrate that this dissimilarity can reflect the chemistry of the elements. Finally, our models were applied to the discovery of new structures in the well-known magnetic material Nd$_2$Fe$_{14}$B. Our models propose 108 new structures, 71 of which are confirmed to converge to local-minimum-energy structures with formation energy less than 0.1 eV by first-principles calculations.
			
		\end{abstract}
		
		
	\end{frontmatter}
	
	%\linenumbers
	
	\section{Introduction}
	Development of novel materials for modern applications is a central focus of materials science. Considerable research effort has been expended to develop novel material systems with desirable properties. Researchers generally utilize their knowledge of physics and/or chemistry, through a set of chemical/physical rules, to guide their search for certain types of materials. However, the diversity of material systems is often not represented by such simple sets of chemical/physical laws, especially for lanthanide and transition-metal compounds. In contrast, recent developments in computational and experimental materials data and advances in computing techniques have allowed machine learning algorithms, especially deep learning techniques, to represent hidden complex chemical/physical concepts \cite{ceder2,ofm,ofm1,ceder2,ceder1,AB_perovskite,Materials_cartography_chem_mater_2015,JCP_LMM,JCP_parallel_lasso_SMM,Pham:yu5018,BehlerJCP,behlerPRL,data_mining_materials_science_PRB2012}. In the last decade, materials research has been marked by the emergence of a new research field called ``materials informatics," which focuses on the use of machine learning algorithms to represent or extract hidden chemistry/physics and to use that for the development of new materials.
	
	Typically, the search for a new material involves exploration of the phase space or potential energy surface (PES) to discover minimum-energy atomic configurations \cite{random_search_1,random_search_2,random_search_3,basin_hopping,uspex_1,uspex_2,uspex,CALYPSO1,CALYPSO2}. However, an exhaustive search of all the possibilities is computationally expensive and, based on current computing performance, is almost impossible. Historically, chemists have attempted to summarize chemical knowledge as chemical rules for the local structures of molecules and solids. For instance, the octet rule has been widely used for the determination of the structure of organic molecules. However, the diversity of chemistry cannot simply be represented by a small set of heuristic rules. Therefore, an understanding of the hidden chemistry/physics of materials is highly desirable so as to capture their diversity, and identify novel materials with new atomic configurations for application in modern industries. Application of advanced machine learning algorithms to materials science using open accessible datasets \cite{OQMD,OQMD1,materialsAPI,convex_hull,mp,MaterialsProject,ceder1} is also expected to accelerate the development of new materials for modern technologies. 
	
	From fundamental chemistry, one can learn that although the chemistry of the transition metals and rare-earth elements is rich, they only exist in a limited number of local chemical environments. For instance, Fe, in most of its compounds, prefers an octahedral structure. In this study, we propose a novel method to identify preferred chemical environments of an atom using material data, in which we develop models to recognize the preferred structures for an atom, i.e., the models for predicting the possibility that an atom can be paired with a local chemical environment in materials. To build the models, we  employ one-hot vectors corresponding to the valence electron configuration to represent atoms and use the Voronoi tessellation method \cite{ofm,ofm1} to determine the chemical environment of atoms in the material. Two multilayer perceptrons are utilized to convert the information from center atoms and their environments to extract the embedding features. These features are then concatenated to form the local-structure feature vectors (i.e., feature vectors for a cluster of atoms formed by a center atom and its neighboring atoms). Next, we implement another multilayer perceptron to map the local-structure feature vector into a number which represent the possibility by the ''Sigmoid' function, as shown in Fig. \ref{fig:model}. 
	
	The models are then utilized to construct recommendation systems that suggest a list of atoms for a given environment in a structure. New structures were then generated by replacing the center atom of the original structure by the recommend atoms. 
	Based on the recommended atoms, we propose a novel approach for estimating the chemical dissimilarity between elements, and using a hierarchical clustering analysis and visualization with the multidimensional scaling (MDS) algorithm, demonstrate that this dissimilarity can capture the chemistry of the elements. In particular, the separation of lanthanide elements from transition metals can be achieved through hierarchical clustering, and distinct patterns for the $3d$, $4d$, and $5d$ transition metals in the latent space can be obtained through MDS. Finally, through single substitutions, the models are employed to predict new materials based on one of the most important magnet materials, Nd$_2$Fe$_{14}$B, which was discovered by Sagawa et al. \cite{doi:10.1063/1.333572}. We employ a model ensemble to suggest 108 new materials, 71 of which have relatively low formation energy (less than 0.1 eV/atom). Note that our approach requires no expert labeling process and is purely data-driven for generalizing the hidden chemistry/physics of materials. 
 The remainder of this paper is organized as follows: In Section 2, we describe the local structure representation, the models for learning the likelihood of structures, and the data used for model training; in Section 3, we present the main results and discuss their implications; and finally, in Section 4, we summarize the study.

	\section{Methodology}
	\subsection{Local structure representation}
	The key aim of this work was to develop machine learning models that can generalize chemical structure information to predict stable structures of materials. We focused on the chemical structures and environment of an atom in a solid material, and encoded those structures using feature vectors conveying the chemistry and symmetry information. Previously, to facilitate application of machine learning algorithms for the mining of hidden knowledge from materials datasets, we developed a novel descriptor called the orbital field matrix (OFM) for the representation of local structures in solids \cite{ofm,ofm1,Pham:yu5018,doi:10.1063/5.0015977}, which utilized the one-hot vector of valence electrons and the Voronoi analysis of atomistic structures of solids. In current work, we also designed feature vectors of atoms by employing one-hot encoding to represent the valence electron configurations, $\vec{O}_{a}$, using a dictionary comprising the valence subshell orbitals: $\{s^1, s^2, p^1, p^2, ..., p^6, d^1, d^2, ..., d^{10}, f^1, f^2, ..., f^{14}\}$. Here, the superscripts indicate the numbers of electrons in the subshells. For example, the valence electron configuration of Fe ($3d^64s^2$) is represented by a 32-element vector with all but the $2^{nd}$ and $14^{th}$ elements being zero; these elements correspond to the $s^2$ and $d^6$ subshells, respectively, and are set to one. 
	
	As suggested by O'Keeffe \cite{Okeeffe_coordination_number}, the chemical environment of an atom was determined by its neighboring atoms which were defined by Voronoi analysis \cite{ofm}. In particular, using a cutoff radius, we determined the set of atoms including the center atom and its neighborhood. Next, the Voronoi polyhedron associated with the center atom was determined by performing 3D Voronoi analysis on this set. The neighboring atoms were then defined as the ones sharing the same faces as the center atom. The neighboring atom information obtained from this exercise was considered as the chemical environment of the center atom. A weighted sum of the one-hot vectors of all the neighboring atoms was done to obtain the chemical environment feature vector ($\vec{O}_e$), i.e.,
	\begin{equation}
		\vec{O}_e = \sum_k w_k O_a ^{(k)}\mbox{,}
	\end{equation}
	where $k$ is the index of the neighboring atoms, $ O_a ^{(k)}$ is the one-hot vector of the $k^{th}$ neighbor atom, and $w_k$ is the weight associated with the neighbor atom. Intuitively, the solid angles formed by the center atom, and the faces of its Voronoi polyhedron conveys information on the interaction between the center atom and its neighboring atoms. Therefore, the weights were defined as a function of the solid angles and the distances between the center atom and the corresponding neighboring atom, i.e., $w_k = w_k(\theta_k, r_k)$, where $\theta_k$ and $r_k$ are the solid angles and distances of the neighboring atoms to the central atom, respectively. In this work, we aimed to use only the valence electron configuration (chemistry) of the atoms in a chemical environments and the local structure geometry (i.e., the shapes of the Voronoi polyhedra). As an example, this treatment would imply that the local structure of a cubic KBr crystal, with each K (at the center) surrounded by six Br atoms, is identical to the structure of NaCl, with Na at the center and six surrounding Cl atoms. Thus, we ignored $r_k$ and used $w_k = \frac{\theta_k}{\theta_{max}}$, where $\theta_{max}$ is the maximum solid angle between the central atom and the faces of its Voronoi polyhedron. We defined the environment feature vector as follows:
	\begin{equation}
		\label{eq:env}
		\vec{O}_e = \sum_k \frac{\theta_k}{\theta_{max}}O_a ^{(k)} \mbox{.}
	\end{equation}
	We again emphasize that our definition of the chemical environment can capture chemistry with valence electrons of atoms and symmetry by Voronoi polyhedra.	
	
	\subsection{Modeling the local structure likelihood}
	
	We employed deep neural networks to investigate the possibility of a local structure being observed in a material; this possibility was termed as the local structure likelihood. In other words, we developed deep learning models to generalize the hidden chemistry of solids to identify types of local structures for an atom. The proposed scheme of the model is shown in Fig. \ref{fig:model}. In this model, the primitive information of an atom and its environment (i.e., its input feature vector) was transformed into a latent space with predefined dimensions using independent feature extractors which were implemented by two multilayer perceptrons (the green and blue parts, respectively). The one-hot vector encoding the valence-electron configuration of the center atom were the input vectors, and the feature vectors defined by Eq. \ref{eq:env} were the chemical environment inputs.  The embedding feature vectors of the latent space were produced after the atom and environment input vectors were passed through the feature extractors. These embedding feature vectors of the atom and its environment were concatenated to create feature vectors for the local structure. Finally, the local-structure feature vectors were used as inputs for a multilayer perceptron with one output neuron with the Sigmoid activation, which provided the local structure likelihood. Since this model describes the local structure likelihood using the OFM, this model has been termed DeepOFM.
	
	\begin{figure}[!h]
		\centering
		\includegraphics[width=0.48\textwidth]{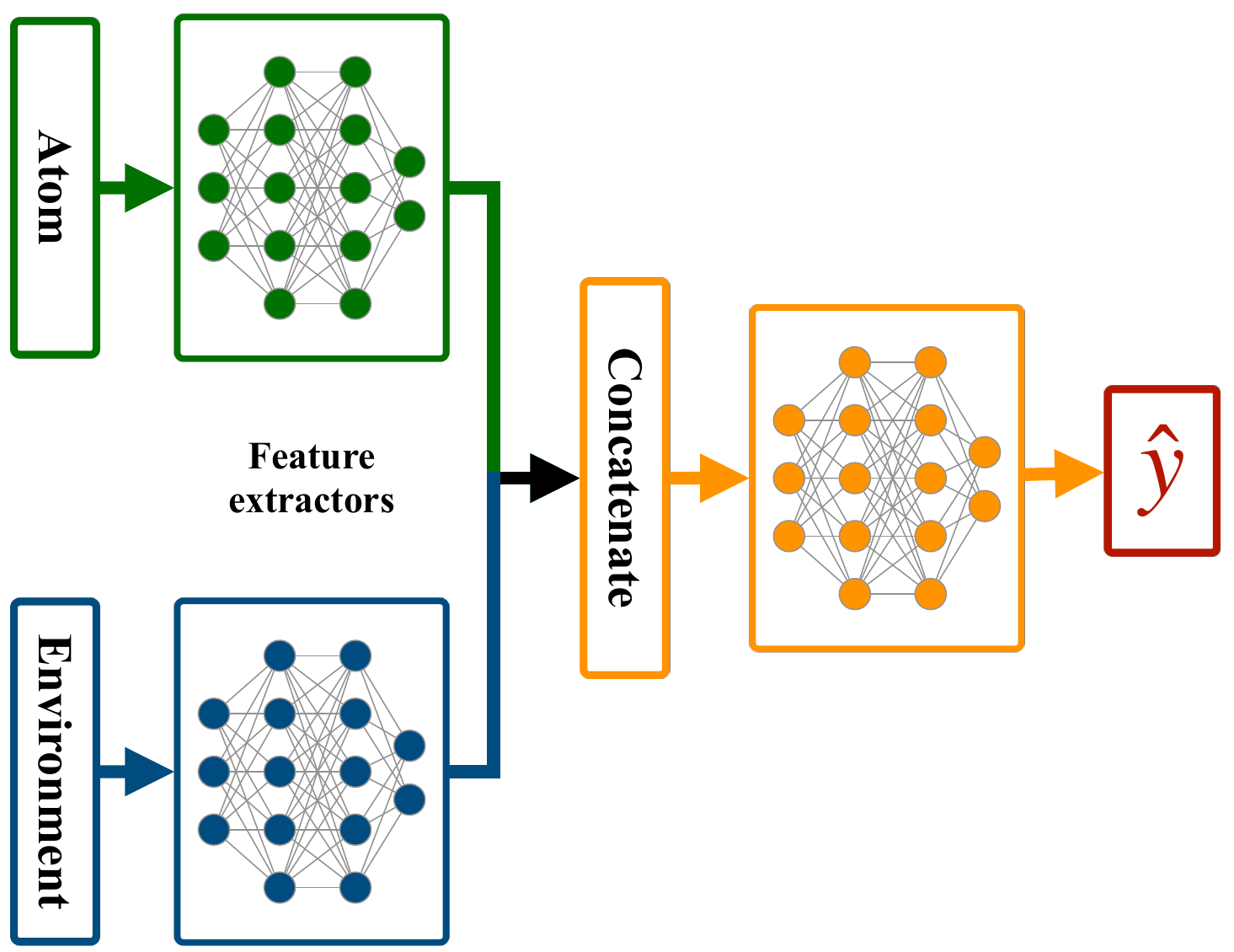}
		\caption{\label{fig:model.pdf} Model for learning local structure likelihood: Embedding feature extraction for environments and center atoms (blue and green) and local structure likelihood estimation (yellow), with information flow indicated by arrows (tensors).}
	\end{figure}
	
\begin{figure}[!h]
		\centering
		\includegraphics[width=0.48\textwidth]{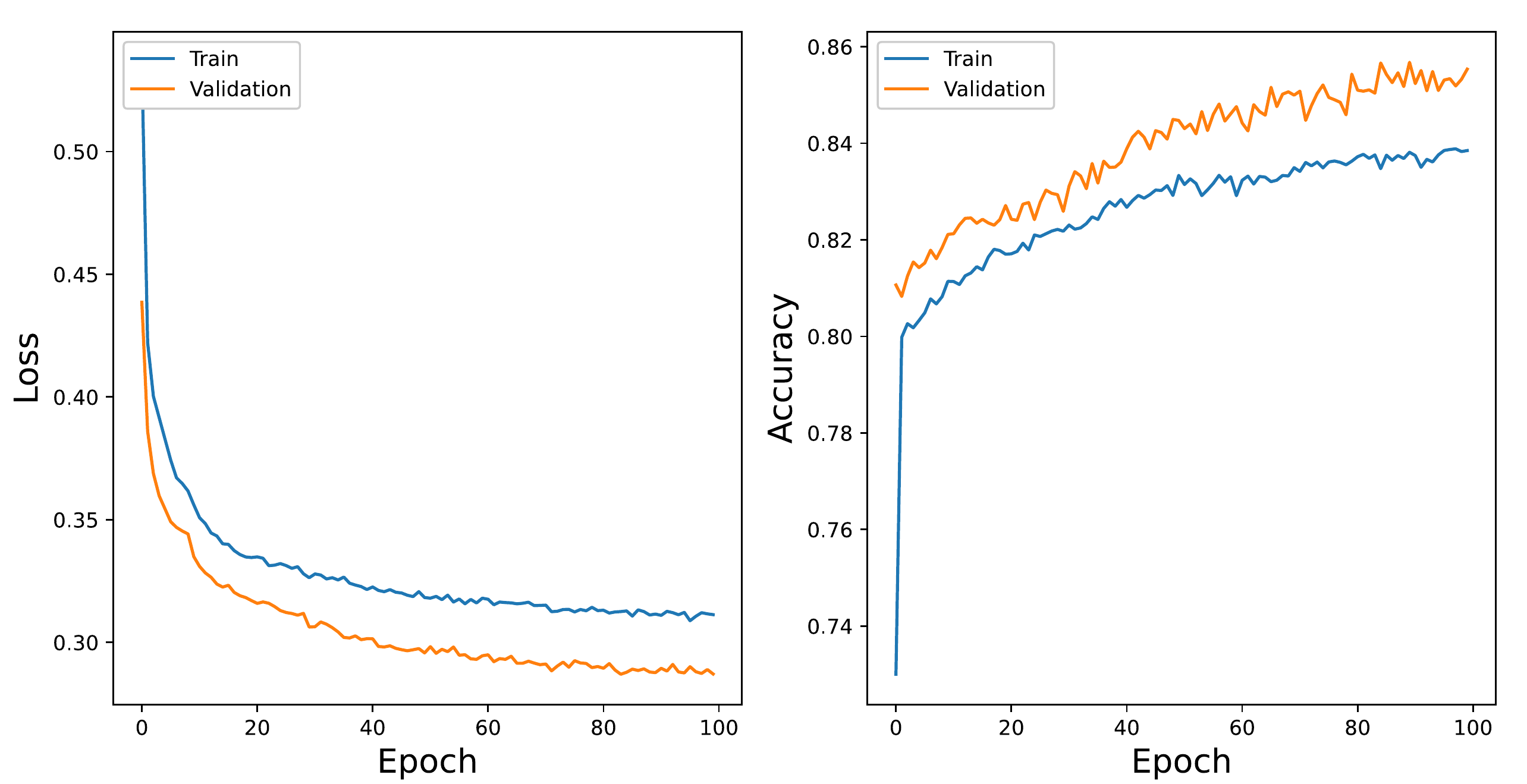}
		\caption{\label{fig:learning_curve} Summary of learning curves obtained for 32 embedding dimensions and Sigmoid activation.}
	\end{figure}

	 In the present study, we examined embedding features with dimensions of 4, 8, 16, or 32, i.e., the output neuron number was 4, 8, 16, or 32, respectively. As shown in Fig. \ref{fig:model}, the atom and environment embedding features were concatenated to form the feature vectors of the corresponding pairs. These vectors were fed to a two-layer perceptron with 128 neurons in each layer. Then, the information, i.e., the final-layer output, was passed to a sigmoid output that represented the pair likelihood. In this model, to represent the local structure likelihood, we investigated various activation functions for the feature extractors and network, including ``rectified linear unit’’ (``ReLU’’) \cite{Schmidhuber_2015}, ``tangent hyperbolic’’ (``Tanh’’), and ``Sigmoid.’’ The binary cross-entropy loss function and adaptive gradient (ADAGRAD) optimizer \cite{adagrad,Schmidhuber_2015}, with a batch size of 32, were employed for model training, where the model was implemented over 100 epochs using the Tensorflow/Keras library \cite{chollet2015keras}. Fig. \ref{fig:learning_curve} shows typical learning curves for our model, which indicate good learning behavior.
	
	\subsection{Data}
	We obtained the structures relaxed using calculations based on density functional theory (DFT), which were provided by the Open Quantum Materials Database (OQMD) \cite{OQMD} for formation energy. We consider this data of structures as the ground truth and apply the data mining approach with deep learning models to generalize the hidden chemistry/physics for identifying the preferred local structure for atoms. We focused on the search for new combinations of lanthanide transition-metal compounds for new permanent magnets.
	To construct a training dataset for our deep neural network, we collected data from the OQMD \cite{OQMD} Repository Version 1. We queried compounds consisting of (1) two transition metals (TT--bimetal), (2) lanthanide atoms and transition-metal atoms (LAT--bimetal), (3) LAT and light (X) atoms, and (4) TT and X atoms. The following sets of transition metals, lanthanides, and X elements were respectively used for this purpose: $\{$Sc, Ti, V, Cr, Mn, Fe, Co, Ni, Cu, Zn, Y, Zr, Nb, Mo, Tc, Ru, Rh, Pd, Ag, Cd, Hf, Ta, W, Re, Os, Ir, Pt, Au$\}$; $\{$La, Ce, Pr, Nd, Pm, Sm, Eu, Gd, Tb, Dy, Ho, Er, Tm, Yb, Lu$\}$; and \{B, C, N, O\}. Based on this, we obtained a dataset of 4220 compounds with 1510 LATX, 1311 TTX, 692 LAT, and 707 TT compounds. From this dataset, we obtained 24494 unique local structures, which were used to train the models to predict local structure likelihood. Many of the local structures that were symmetry-equivalent, were removed by comparing the environment feature vectors. Local structures having a difference of less than 0.001 (Euclidean distance) and the same center atom were considered to be identical. 
		
	\section{Results and discussion}
	\subsection{Learning hidden chemistry}
	
	To train our models, we collected a dataset including positive and negative examples. There are existed 24494 local structures as positive examples which were collected based on 4200 compounds. The negative examples were local structures that were not found in the structure data. For each environment of positive examples, we randomly selected $n_s$ atoms ($n_s$ being 2, 3, 4, or 5) to generate negative examples. A preliminary examination indicated that $n_s$ = 2 was suitable for training the models. Combining the negative and positive examples, we obtained a dataset of 73482 local structures. We then randomly divided this dataset into a training set (80 \%) and a test set (20 \%). By this procedure, we compiled our models into the classification models for positive and negative examples.
	
	To evaluate our models, we selected the top-k atoms to each environment based on their likelihood (estimated using DeepOFM). We then counted the number of environments whose center atoms were found in the top-k recommended atoms by our models. The recommendation recall ($R_k$) was then calculated using the following equation: $R_k = \frac{n^c_k}{n^\circ}$, where $n^c_k$ and $n^\circ$ are the number of environments for which the center atoms were correctly recommended in the top-k and the number of environments in the test set, respectively. We also employed the positive--negative classification accuracy score to measure the model performance. The results are summarized in Table \ref{tab:top5}. It is apparent that most of our experiment settings had accuracy scores exceeding 85 \%, and we could achieve 60 \% recall using Sigmoid activation and 32 embedding features. The results indicate that DeepOFM can accurately represent the local structure likelihood, and hence we can extract the hidden chemistry of solids from material structure data. To access the knowledge generalized by DeepOFM, we compared the recommended substitutions for each pair of atoms in the dataset of 4200 structures.

	\begin{table}[]
		\caption{\label{tab:top5} Recommendation recalls (\%) for test-set environments and DeepOFM accuracy scores  obtained with ReLU, Sigmoid, and Tanh activations. }
		\centering
		\begin{tabular}{lllllll}
			\hline\hline
			Settings    & Top 1 & Top 2 & Top 3 & Top 4 & Top 5 & Score \\
			\hline
			ReLU-4 & 42.1 & 46.6 & 50.3 & 53.3 & 56.9		&0.864\\
			ReLU-8 & 42.7 & 47.1 & 51.1 & 54.7 & 57.4		&0.869\\
			ReLU-16 & 41.9 & 47.4 & 52.5 & 56.3 & 59.3		&0.871\\
			ReLU-32 & 42.7 & 48.2 & 52.8 & 55.7 & 58.3		&0.864\\
			Sigmoid-4 & 42.4 & 47.3 & 52.1 & 55.1 & 58.5	&0.852\\
			Sigmoid-8 & 39.5 & 47.0 & 51.0 & 54.3 & 57.4	&0.856\\
			Sigmoid-16 & 38.6 & 47.2 & 52.0 & 54.4 & 57.2	&0.853\\
			Sigmoid-32 & 39.4 & 48.7 & 53.9 & 57.3 & 60.5	&0.861\\
			Tanh-4 & 34.6 & 46.4 & 52.1 & 55.8 & 58.5		&0.856\\
			Tanh-8 & 41.9 & 46.8 & 50.1 & 53.5 & 56.5		&0.842\\
			Tanh-16 & 38.0 & 45.4 & 49.1 & 52.2 & 56.0		&0.859\\
			Tanh-32 & 41.8 & 47.3 & 51.3 & 54.0 & 56.6		&0.842\\
			\hline\hline
		\end{tabular}
	\end{table}

	\begin{figure}[!h]
		\centering
		\includegraphics[width=0.44\textwidth]{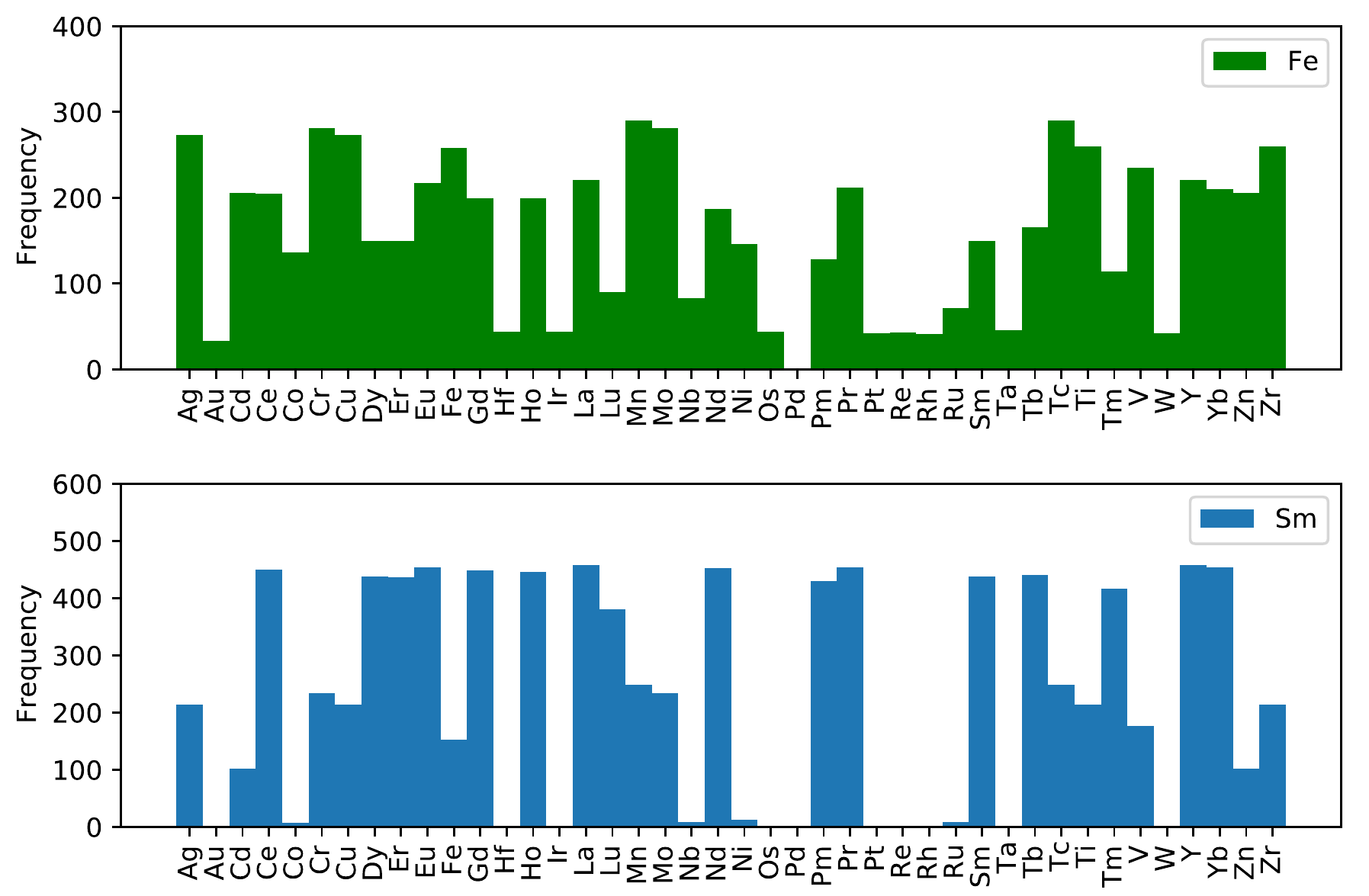}
		\caption{\label{fig:recommend_distribution} Recommendation distributions for Fe (upper) and Sm (lower) obtained by counting the number of recommended substitutions by elements in the reference set for Fe and Sm, using the model with ReLU activation and 32 embedding features for the center atoms and environments.
		}
	\end{figure}
	
	Intuitively, elements that have common chemical properties can be substituted for each other at a higher likelihood. DeepOFM is trained to estimate the local structure likelihoods, and we can utilize these likelihoods to recommend substitutes for a given atom. Hence, by comparing the recommended atoms for atoms in a material dataset, we can access the similarity (or dissimilarity) between elements. Therefore, in this study, we used the collected dataset of 4200 structures as a reference for the dissimilarity measurement. For each element $A$, we collected all environments of $A$ in all materials in the dataset and used DeepOFM to calculate the likelihood of an atom replacing $A$ in each environment. We used a likelihood threshold of 0.6 to select atoms to replace $A$ in these environments and counted the number of times an atom was recommended as a substitute for $A$. Hence, Using a set of elements as reference elements, we could obtain the distribution of number recommendations for $A$ over reference set, which can ben considered as a representation of the chemistry of $A$. In this study, \{Ag Au Cd Ce Co Cr Cu Dy Er Eu Fe Gd Hf Ho Ir La Lu Mn Mo Nb Nd Ni Os Pd Pm Pr Pt Re Rh Ru Sm Ta Tb Tc Ti Tm V W Y Yb Zn Zr\} were used as reference elements. For example, consider Fe and Sm. By pairing all environments of Fe and Sm in the dataset with the reference set, we obtained 16968 and 19782 new local structures, respectively. We employed DeepOFM with ReLU activation and 32 atom and environment embedding features to calculate the likelihoods of the proposed local structures. By eliminating all local structures with likelihoods of less than 0.6, we were left with 3448 and 5787 replacements for Fe and Sm, respectively. By counting the number of instances of a given element among the reference elements, we obtained the distributions of the recommended replacements for Fe and Sm, as shown in Fig. \ref{fig:recommend_distribution}. Clearly, in many cases, an element (such as Ni, Co, or Os) was recommended for the replacement of Fe but not for Sm.  
	
	To quantitatively compare the chemistry of elements (Fe and Sm in this case), we employed the Jensen--Shannon divergence, which is widely used to compare two distributions.
	\begin{figure}[!h]
		\centering
		\includegraphics[width=0.44\textwidth]{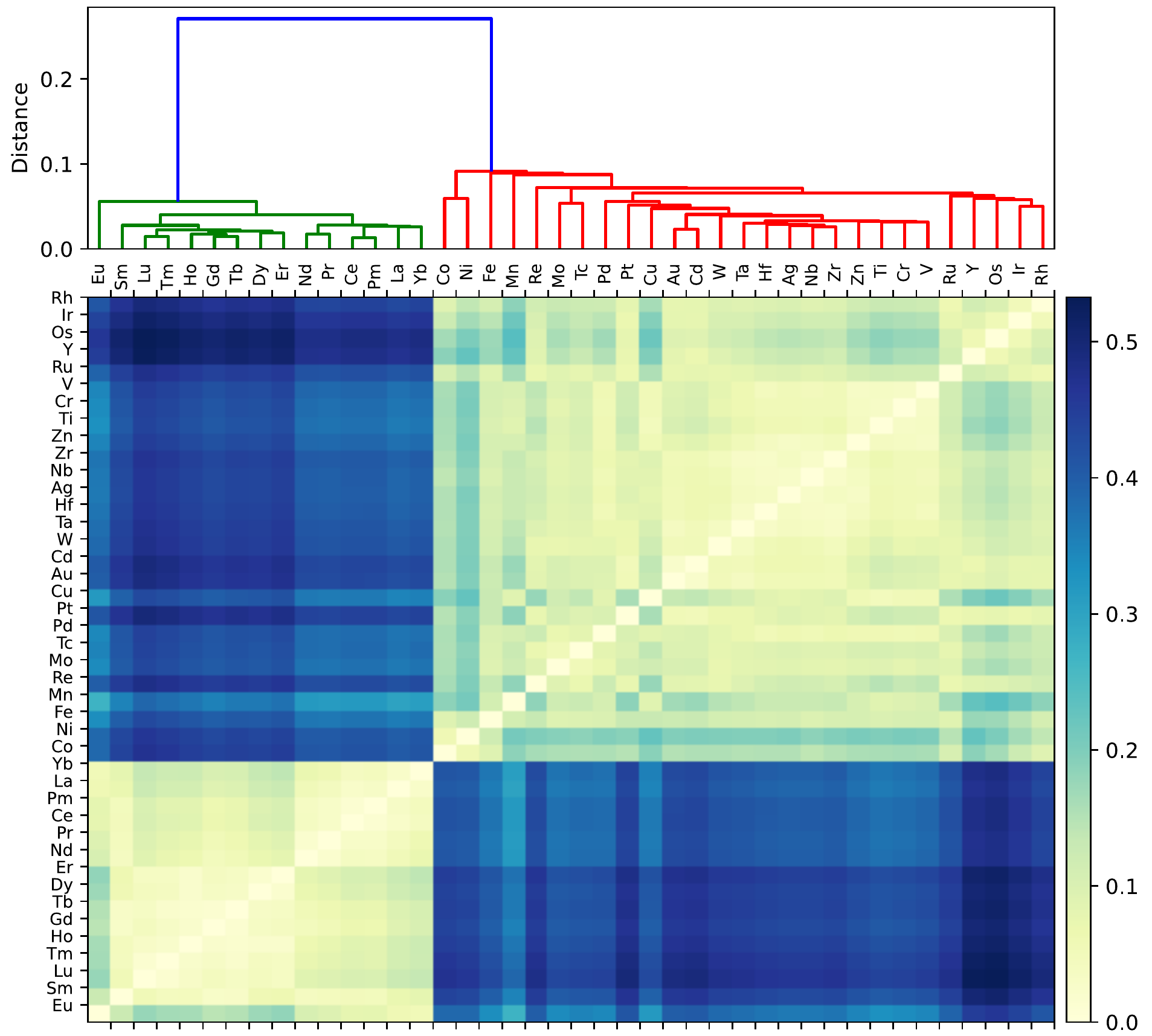}
		\caption{\label{fig:dendogram} Distance matrix and dendrogram of elements obtained using Jensen--Shannon divergence and model with ReLU activation and 32 embedding features for center-atom and environment.}
	\end{figure}
	For instance, we estimated the dissimilarity of elements $A$ and $B$ based on their recommended distributions, $p(i) $ and $q(j)$, respectively, where $i$ and $j$ are elements in the reference set. The Jensen--Shannon divergence was calculated as follows:
	\begin{equation}
		\label{eq:jsd}
		JSD(A||B) = \frac{1}{2} \sum_i p(i)ln\left(\frac{p(i)}{q(i)}  \right) +  \frac{1}{2} \sum_i q(i)ln\left(\frac{q(i)}{p(i)}  \right) \mbox{,}
	\end{equation}
	where $JSD(A||B)$ is the Jensen--Shannon divergence of the distribution of A and B. Note that the Jensen--Shannon divergence is a measurement of the difference in the distribution of $A$ and $B$, i.e., it is an assessment of the chemical dissimilarity of $A$ and $B$. By using Eq. \ref{eq:jsd} and the distributions obtained from the model with ReLU activation and 32 embedding features, we estimated the dissimilarity of Co and Fe as 0.092 and that of Fe and Sm as 0.399. Since both Co and Fe are $3d$ elements, their properties are similar, whereas the properties of Fe and Sm are quite different. Therefore, it can be concluded that the dissimilarity measurements are in good agreement with the underlying chemistry of the elements. This result indicates that our DeepOFM model can learn hidden chemistry from materials data.
	
	Using the Jensen--Shannon divergence, we obtained a dissimilarity matrix for all elements in the reference set, which was used to measure the chemical dissimilarities of the elements. Using this dissimilarity matrix, we applied hierarchical cluster analysis to the elements. The upper panel of Fig. \ref{fig:dendogram} shows the \textit{dendrogram} obtained from the hierarchical cluster analysis, which conveys the hierarchical relationship between groups of elements. Within the hierarchical clustering, we defined a group of elements based on their dissimilarities. For instance, for the dendrogram shown in Fig. \ref{fig:dendogram}, a dissimilarity threshold of 0.15 divides the reference set into two groups, as indicated by the red and green lines. Surprisingly, using this method, the lanthanides could be separated from the rest of the transition-metal elements of the reference set. This result provides strong evidence that the DeepOFM model was able to learn the chemistry of the elements in our dataset, and that our dissimilarity measurement provides a good assessment of the chemical differences between the elements. 
	
	\begin{figure}[!h]
		\centering
		\includegraphics[width=0.44\textwidth]{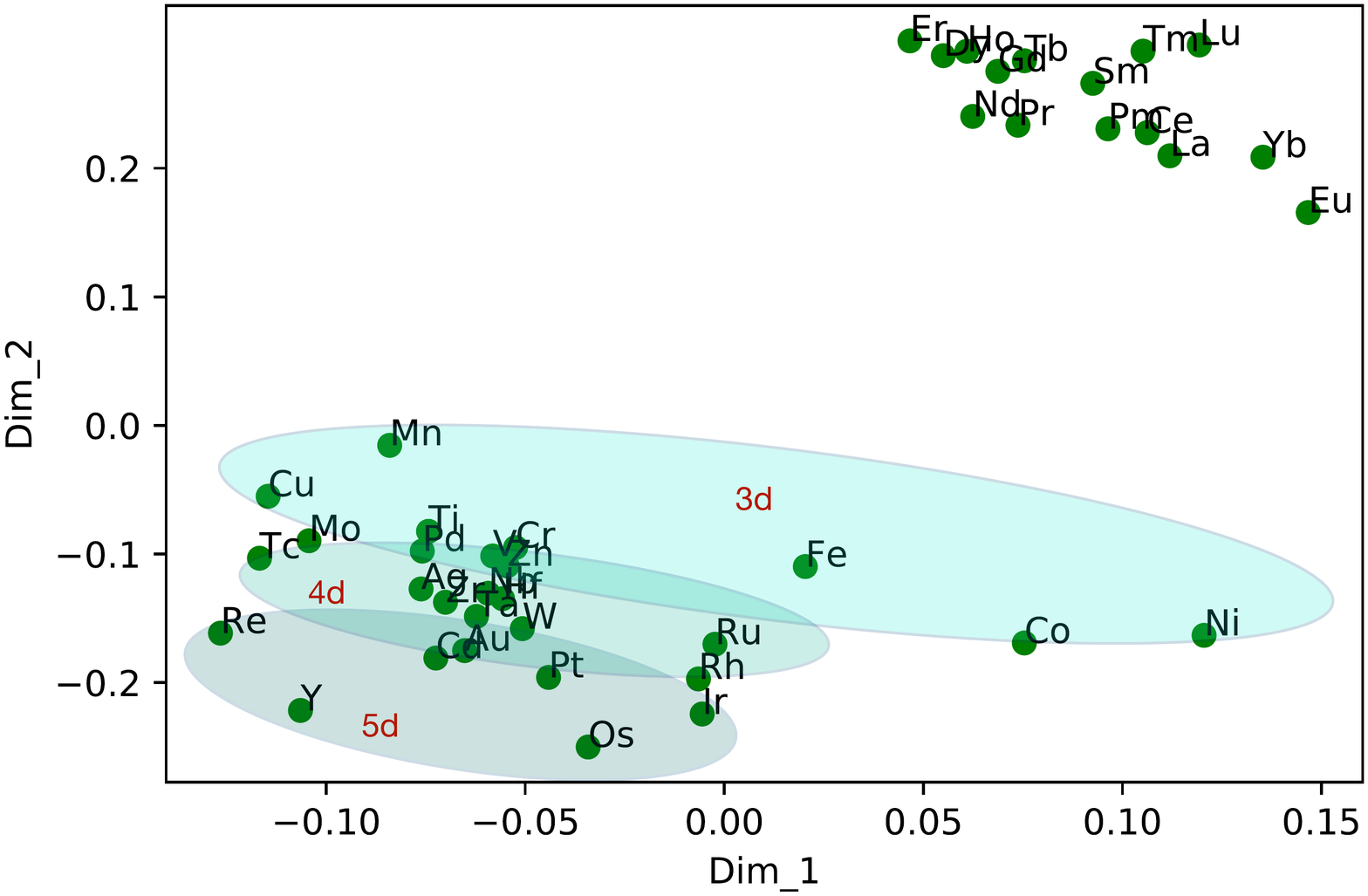}
		\caption{\label{fig:mds} Scatter plot of elements in reference set in latent 2D-space obtained through MDS using distance matrix calculated with Jensen--Shannon divergence.}
	\end{figure}
	
	This result motivated us to learn the element embedding features based on this dissimilarity measurement. We utilized the MDS manifold learning algorithm \cite{model_multivariates}, which extracts latent features by conserving the distance (dissimilarity) between data points. By keeping two dimensions immersed in the reference dataset with our dissimilarity measurement, we obtained element scatter plots as depicted in Fig. \ref{fig:mds}. Again, a clear separation between the lanthanide elements (upper-right) and the rest of the transition metals (lower-left) is apparent. Interestingly, a separation between the $3d$, $4d$, and $5d$ transition-metals is also apparent, as seen in Fig. \ref{fig:mds}.
	
	\subsection{Prediction of new materials}
\begin{figure}[!h]
		\centering
		\includegraphics[width=0.44\textwidth]{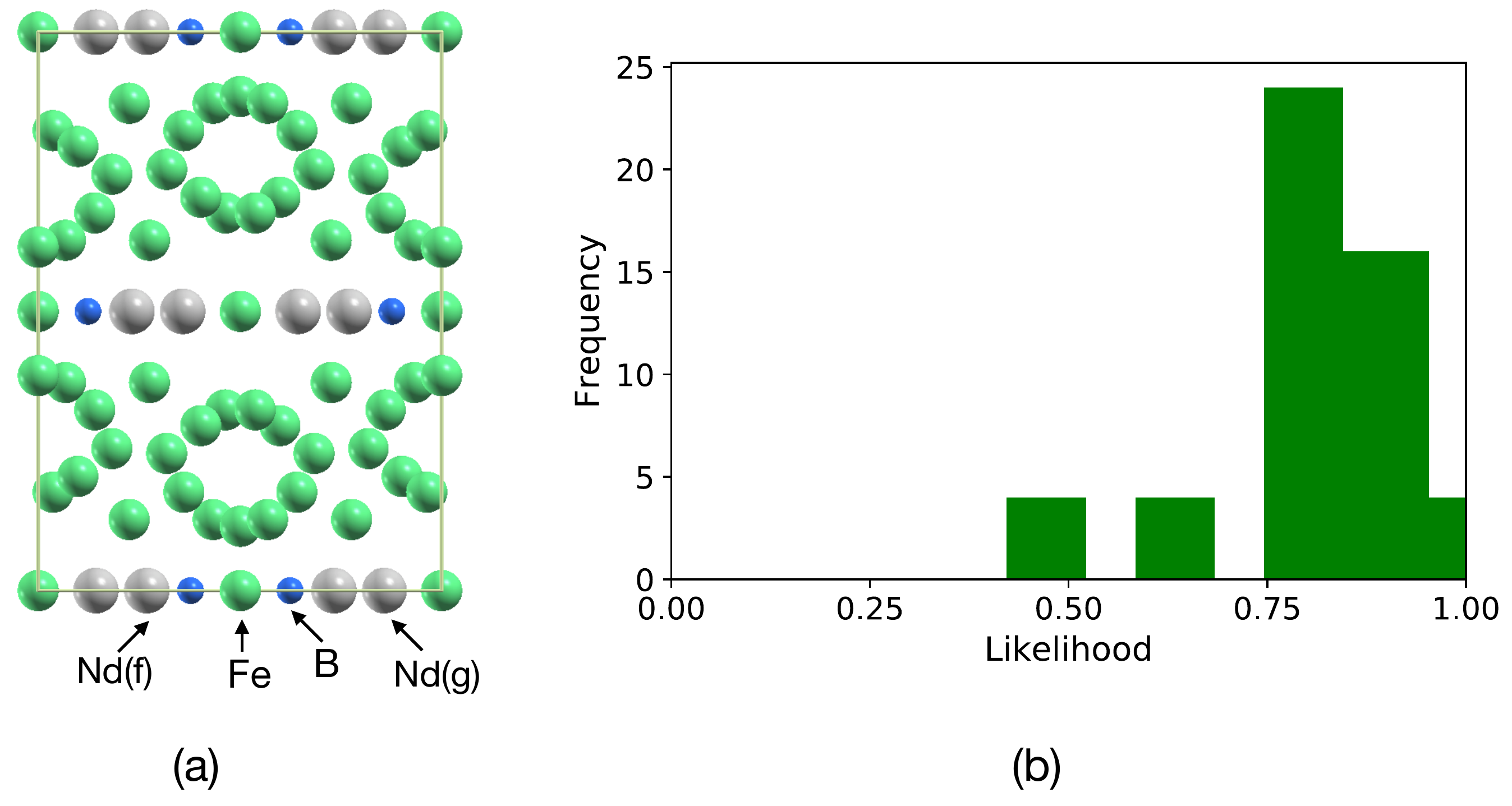}
		\caption{\label{fig:hist_likelihood_68} Supercell of Nd$_2$Fe$_{14}$B (a), and  likelihood distribution of 68 local structures in Nd$_2$Fe$_{14}$B calculated using ReLU activation function with 32 embedding features for both environments and center atoms.}
	\end{figure}

	We focused on predicting new magnetic materials consisting of rare-earth elements, transition metals, and light elements. We adopted Nd$_2$Fe$_{14}$B, the most important magnetic material, as the host structure for the development of these new materials, such that its atoms would be replaced with the target elements. The tetragonal crystal structure of Nd$_2$Fe$_{14}$B consisting of 68 atoms and with a space group of $P4_2/mnm$ was obtained from the Materials Project repository \cite{mp}, as shown in Fig. \ref{fig:hist_likelihood_68}(a). We first examined the DeepOFM models using the parameters provided in Table \ref{tab:top5}, and estimated the likelihood of 68 local structures in the Nd$_2$Fe$_{14}$B supercell. The likelihood distribution of the 68 local structures of Nd$_2$Fe$_{14}$B is depicted in Fig. \ref{fig:hist_likelihood_68}(b). The results show that almost all the 68 local structures had likelihoods exceeding 0.5. This result again indicates that our models can reasonably predict the possibility of local structures observed in a material, based on the generalization of hidden chemistry/physics from material structure data.
	
	We next utilized the DeepOFM models to predict new materials by replacing atoms in Nd$_{2}$Fe$_{14}$B with the following $R$ or $T$ atoms. We extracted the 68 chemical environments using the method described in Section 2.2 and paired these environments with the one-hot vectors for the $R$ or $T$ atoms. The $T$ atoms were selected from the set \{Sc, Ti, V, Cr, Mn, Co, Ni, Cu, Zn, Y, Zr, Nb, Mo, Tc, Ru, Rh, Pd, Ag, Cd, Hf, Ta, W, Re, Os, Ir, Pt, Au\}, and the $R$ atoms were selected from the set \{La, Ce, Pr, Pm, Sm, Eu, Gd, Tb, Dy, Ho, Er, Tm, Yb, Lu\}. By pairing each of the $T$ and $R$ elements as the center atom with the 68 chemical environments, we obtained 2788 environment and center-atom pairs (i.e., hypothesized local structures). We used the feature vectors defined in Eq. \ref{eq:env} to represent the 68 chemical environments, one-hot vectors for the center atoms, and DeepOFM to estimate the likelihood of the local structures.
	
\begin{figure}[!h]
		\includegraphics[width=0.48\textwidth]{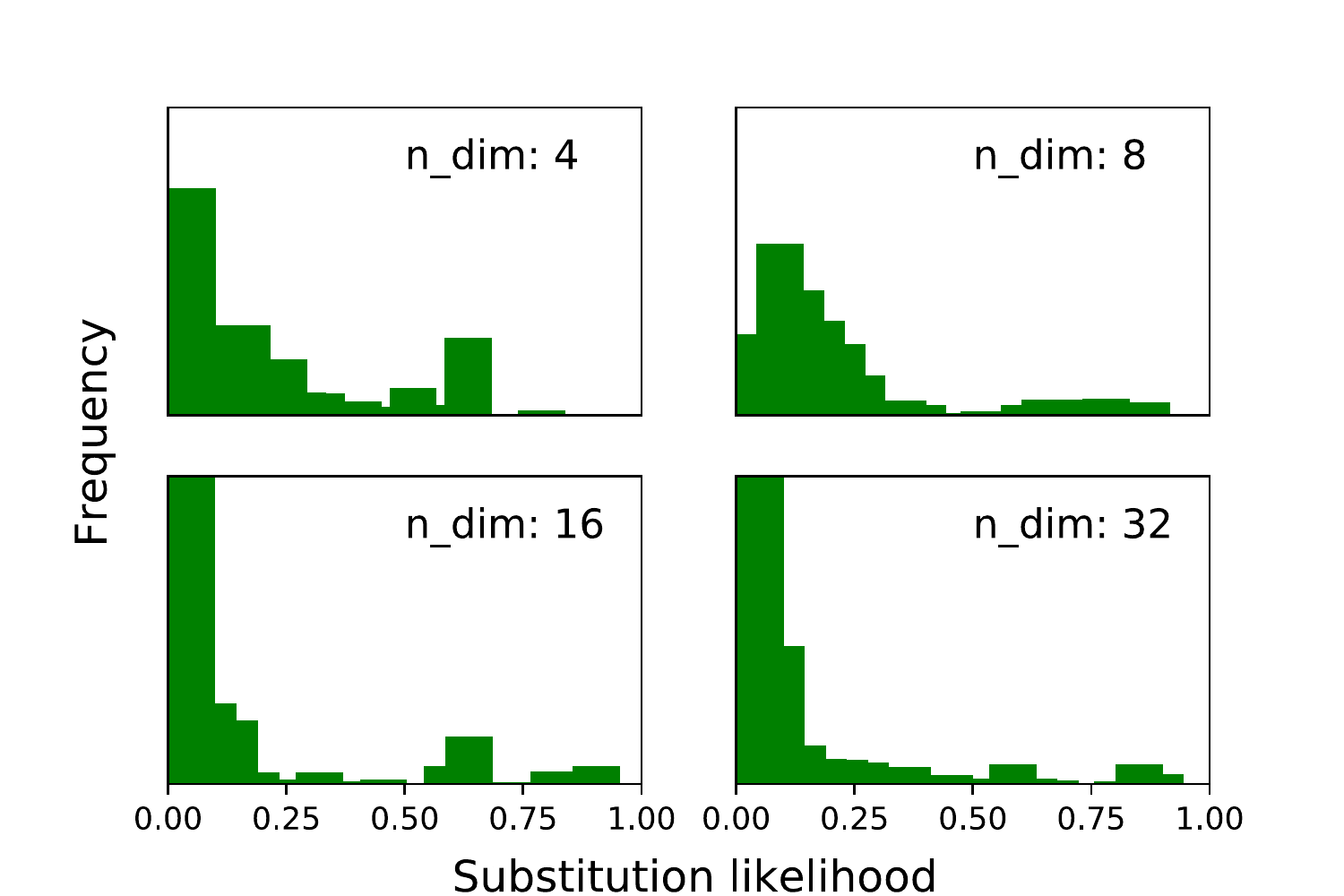}
		\caption{\label{fig:hist_likelihood} Likelihood distributions for proposed local structures obtained by replacing center atoms with $T$ or $R$ atoms using 4, 8, 16, and 32 embedding features.}
	\end{figure}

	Fig. \ref{fig:hist_likelihood} depicts histograms of the local structure likelihood obtained using ReLU activation and 4, 8, 16, and 32 embedding features for both the center atoms and chemical environments. The histograms clearly show that the likelihood of a large number of local structures was below 0.5, and thus only a small number of local structures were recommended by the DeepOFM models. We used the threshold of 0.5 to eliminate unlikely substitutions for Nd$_2$Fe$_{14}$B. Table \ref{table:prediction} lists the number of suggested local structures using the ReLU, Sigmoid, and Tanh activations and 4, 8, 16, and 32 embedding features, respectively. For 32 embedding features, the DeepOFM model predicted 252, 292, and 148 substitutions with the ReLU, Sigmoid, and Tanh activations, respectively.
	
	To improve the confidence of the recommendations, we only used the structures generated by replacing one atom in Nd$_2$Fe$_{14}$B with the atom proposed by all models in a selected set of models. We employed models using the ReLU activation to select new structures. Thus, the selected structures were those recommended by the agreement of all models employing 4, 8, 16, and 32 embedding features. Based on this procedure, we obtained 108 new structures. By examining the original atoms in the host structure and the recommended substitutes, we found that Nd can be replaced with Sc, Y, Sm, Yb, La, Eu, Gd, Dy, Ce, Pr, or Er, whereas Fe could only be replaced with Ni and Co. This observation indicates that our models predicted the subsititutes with highly similar elements.
	
	\begin{table}[!h]
		\caption{\label{table:prediction}Number of recommended substitutions obtained using ReLU, Sigmoid, and Tanh activation functions with embedding feature numbers of 4, 8, 16 and 32.}
		\centering
		\begin{tabular}{p{2cm}p{1cm}p{1cm}p{1cm}p{1cm}}
			\hline\hline	
			& 4   & 8   & 16  & 32  \\
			\hline
			ReLU    & 388 & 364 & 428 & 252 \\
			Sigmoid & 452 & 448 & 252 & 292 \\
			Tanh    & 468 & 744 & 276 & 148 \\
			\hline\hline
		\end{tabular}
	\end{table}
	
	Finally, we performed DFT simulations to evaluate the stability of the 108 recommended structures. We adopted the DFT settings of OQMD \cite{OQMD,OQMD1} to relax the suggested structures and evaluate the local-minimum-energy structures: ``coarse relaxation’’ followed by ``fine relaxation’’ and ``standard’’ calculations. The stability of the structures were investigated by evaluating the formation energies owing to the substitutions, which are calculated as follows: $\Delta E_f = E_{subs} + \mu_S - (E_{Nd_2Fe_{14}B} + \mu_T)$,
	where $E_{subs}$ and $E_{Nd_2Fe_{14}B}$ are the total energies per unit cell of the substitutions and the pristine $Nd_2Fe_{14}B$, respectively, and $\mu_S$ and $\mu_T$ are the chemical potentials of the source atom (S-atom) and the target atom (T-atom), respectively. The chemical potentials were calculated as the total energies per atom of the ground-state structures of the $S$ and $T$ atoms. The ground states were selected from the OQMD database \cite{OQMD,OQMD1}. For example, the ground-state structures of Fe, Nd, Co, and B are BCC(Im-3m), HCP(P6$_3$/mmc), FCC(Fm-3m), and $\alpha$-boron (R-3m), respectively.

	The calculations reveal 71 substituted structures with formation energies less than 0.10 eV/atom. For the case of Fe, we observed that Co-substitution had the mean of formation energy of -0.19 eV/atom, and that of Ni-substitution was -0.10 eV/atom. These results agree with the measurement of dissimilarity by Eq. \ref{eq:jsd} wherein $JSD(Fe||Co)=0.092$ is slightly smaller than $JSD(Fe||Ni)=0.119$. These results are also an indication of the accurate recommendations by DeepOFM, with a relatively low substitution formation energy implying a small chemical dissimilarity between the studied elements. For the case of Nd, two crystallographically equivalent sites could be identified, which are labeled as Nd(f) and Nd(g) in Fig. \ref{fig:hist_likelihood_68} (a).  The Nd(g) sites were observed to be favorable to substitution by Dy, Gd, Er, Sm, Sc, Y with the formation energies of substitutions less than 0.1 eV, whereas only Sc and Y were observed to substitute the Nd(f) sites. 

	\section{Conclusions}
	A novel method for learning hidden knowledge from materials data was presented, in which a DeepOFM model was used to assess the likelihood of a local structure being stable in a material. In this approach, neural networks are used to determine the local structure likelihoods. The developed models recommend a list of replacement atoms for a given environment in a structure. Accordingly, new structures are generated by replacing the atoms in the original structure with the recommended ones. 
	Based on these recommendations, we performed a dissimilarity measurement between the elements. This measured dissimilarity are shown to reflect the chemistry of the elements through a hierarchical clustering analysis and latent feature extraction with the MDS algorithm.
	We used our models to determine substitution atoms for Nd$_2$Fe$_{14}$B to create new materials. Our models proposed 108 new materials, and the DFT calculations also confirmed that 71 of the recommended substitutions have relatively low formation energy.  
	
	\section*{Acknowledgments}
	This research was funded by the Vietnam National Foundation for Science and Technology Development (NAFOSTED) under grant number 103.01-2019.30. We are thankful to Dr. Pham Van Hai (Center for Computational Science (CCS) and Faculty of Physics, Hanoi National University of Education) for their valuable support with the VASP calculations.  
	
	\section*{References}
	\bibliography{main}

\begin{thebibliography}{10}
\expandafter\ifx\csname url\endcsname\relax
  \def\url#1{\texttt{#1}}\fi
\expandafter\ifx\csname urlprefix\endcsname\relax\def\urlprefix{URL }\fi
\expandafter\ifx\csname href\endcsname\relax
  \def\href#1#2{#2} \def\path#1{#1}\fi

\bibitem{ceder2}
G.~Hautier, C.~Fischer, V.~Ehrlacher, A.~Jain, G.~Ceder, Data mined ionic
  substitutions for the discovery of new compounds, Inorganic Chemistry 50~(2)
  (2011) 656--663.

\bibitem{ofm}
T.~Lam~Pham, H.~Kino, K.~Terakura, T.~Miyake, K.~Tsuda, I.~Takigawa,
  H.~Chi~Dam, Machine learning reveals orbital interaction in materials, Sci
  Technol Adv Mater 18~(1) (2017) 756--765.

\bibitem{ofm1}
T.-L. Pham, N.-D. Nguyen, V.-D. Nguyen, H.~Kino, T.~Miyake, H.-C. Dam, Learning
  structure-property relationship in crystalline materials: A study of
  lanthanide–transition metal alloys, J. Chem. Phys 148~(20) (2018) 204106.

\bibitem{ceder1}
L.~Yang, G.~Ceder, Data-mined similarity function between material
  compositions, Phys. Rev. B 88 (2013) 224107.

\bibitem{AB_perovskite}
P.~V. Balachandran, A.~A. Emery, J.~E. Gubernatis, T.~Lookman, C.~Wolverton,
  A.~Zunger, Predictions of new $\mathit{AB}{\mathrm{o}}_{3}$ perovskite
  compounds by combining machine learning and density functional theory, Phys.
  Rev. Materials 2 (2018) 043802.

\bibitem{Materials_cartography_chem_mater_2015}
O.~Isayev, D.~Fourches, E.~N. Muratov, C.~Oses, K.~Rasch, A.~Tropsha,
  S.~Curtarolo, {Materials Cartography: Representing and Mining Materials Space
  Using Structural and Electronic Fingerprints}, Chem. Mater 27 (2015) 735.

\bibitem{JCP_LMM}
T.~L. Pham, H.~Kino, K.~Terakura, T.~Miyake, H.~C. Dam, Novel mixture model for
  the representation of potential energy surfaces, J. Chem. Phys. 145~(15)
  (2016) 154103.

\bibitem{JCP_parallel_lasso_SMM}
H.~C. Dam, T.~L. Pham, T.~B. Ho, A.~T. Nguyen, V.~C. Nguyen, Data mining for
  materials design: A computational study of single molecule magnet, J. Chem.
  Phys. 140~(4) (2014) 044101.

\bibitem{Pham:yu5018}
T.-L. Pham, D.-N. Nguyen, M.-Q. Ha, H.~Kino, T.~Miyake, H.-C. Dam,
  \href{https://doi.org/10.1107/S2052252520010088}{{Explainable machine
  learning for materials discovery: predicting the potentially formable
  Nd{--}Fe{--}B crystal structures and extracting the structure{--}stability
  relationship}}, IUCrJ 7~(6) (2020) 1036--1047.
\newblock \href {http://dx.doi.org/10.1107/S2052252520010088}
  {\path{doi:10.1107/S2052252520010088}}.
\newline\urlprefix\url{https://doi.org/10.1107/S2052252520010088}

\bibitem{BehlerJCP}
J.~Behler, Atom-centered symmetry functions for constructing high-dimensional
  neural network potentials, J. Phys. Chem 134 (2011) 074106.

\bibitem{behlerPRL}
J.~Behler, M.~Parrinello, Generalized neural-network representation of
  high-dimensional potential-energy surfaces, Phys. Rev. Lett. 98 (2007)
  146401.

\bibitem{data_mining_materials_science_PRB2012}
S.~Yousef, G.~Da, N.~Thanh, B.~Scotty, C.~J. R., A.~Wanda, Data mining for
  materials: Computational experiments with $ab$ compounds, Phys. Rev. B 85
  (2012) 104104.

\bibitem{random_search_1}
C.~J. Pickard, R.~J. Needs, High-pressure phases of silane, Phys. Rev. Lett. 97
  (2006) 045504.

\bibitem{random_search_2}
C.~J. Pickard, R.~J. Needs, Structure of phase iii of solid hydrogen, Nature
  Physics 3.

\bibitem{random_search_3}
C.~J. Pickard, R.~J. Needs, Ab initio random structure searching, Journal of
  Physics: Condensed Matter 23~(5) (2011) 053201.

\bibitem{basin_hopping}
K.~John, L.~F.~J. J., N.~M. A., V.~Jacco, Progress in crystal structure
  prediction, Chemistry – A European Journal 17~(38) (2011) 10736--10744.

\bibitem{uspex_1}
A.~R. Oganov, A.~O. Lyakhov, M.~Valle, How evolutionary crystal structure
  prediction works—and why, Accounts of Chemical Research 44~(3) (2011)
  227--237.

\bibitem{uspex_2}
A.~O. Lyakhov, A.~R. Oganov, H.~T. Stokes, Q.~Zhu, New developments in
  evolutionary structure prediction algorithm uspex, Computer Physics
  Communications 184~(4) (2013) 1172 -- 1182.

\bibitem{uspex}
C.~W. Glass, A.~R. Oganov, N.~Hansen, Uspex—evolutionary crystal structure
  prediction, Computer Physics Communications 175~(11) (2006) 713 -- 720.

\bibitem{CALYPSO1}
Y.~Wang, J.~Lv, L.~Zhu, Y.~Ma, Crystal structure prediction via particle-swarm
  optimization, Phys. Rev. B 82 (2010) 094116.

\bibitem{CALYPSO2}
Y.~Zhang, H.~Wang, Y.~Wang, L.~Zhang, Y.~Ma, Computer-assisted inverse design
  of inorganic electrides, Phys. Rev. X 7 (2017) 011017.

\bibitem{OQMD}
J.~E. Saal, S.~Kirklin, M.~Aykol, B.~Meredig, C.~Wolverton, Materials design
  and discovery with high-throughput density functional theory: The open
  quantum materials database (oqmd), JOM 65~(11) (2013) 1501--1509.

\bibitem{OQMD1}
S.~Kirklin, J.~E. Saal, B.~Meredig, A.~Thompson, J.~W. Doak, M.~Aykol,
  S.~R{\"u}hl, C.~Wolverton, The open quantum materials database (oqmd):
  assessing the accuracy of dft formation energies, Npj Computational Materials
  1 (2015) 15010 EP --.

\bibitem{materialsAPI}
S.~P. Ong, S.~Cholia, A.~Jain, M.~Brafman, D.~Gunter, G.~Ceder, K.~A. Persson,
  The materials application programming interface (api): A simple, flexible and
  efficient (api) for materials data based on representational state transfer
  (rest) principles, Comput. Mater. Sci 97 (2015) 209 -- 215.

\bibitem{convex_hull}
W.~Sun, S.~T. Dacek, S.~P. Ong, G.~Hautier, A.~Jain, W.~D. Richards, A.~C.
  Gamst, K.~A. Persson, G.~Ceder, The thermodynamic scale of inorganic
  crystalline metastability, Science Advances 2~(11).

\bibitem{mp}
A.~Jain, S.~P. Ong, G.~Hautier, W.~Chen, W.~D. Richards, S.~Dacek, S.~Cholia,
  D.~Gunter, D.~Skinner, G.~Ceder, K.~a. Persson, {The Materials Project: A
  materials genome approach to accelerating materials innovation}, APL
  Materials 1~(1) (2013) 011002.

\bibitem{MaterialsProject}
A.~Jain, S.~P. Ong, G.~Hautier, W.~Chen, W.~D. Richards, S.~Dacek, S.~Cholia,
  D.~Gunter, D.~Skinner, G.~Ceder, K.~a. Persson, Commentary: The materials
  project: A materials genome approach to accelerating materials innovation,
  APL Materials 1~(1) (2013) 011002.

\bibitem{doi:10.1063/1.333572}
M.~Sagawa, S.~Fujimura, N.~Togawa, H.~Yamamoto, Y.~Matsuura,
  \href{https://doi.org/10.1063/1.333572}{New material for permanent magnets on
  a base of nd and fe (invited)}, Journal of Applied Physics 55~(6) (1984)
  2083--2087.
\newblock \href {http://dx.doi.org/10.1063/1.333572}
  {\path{doi:10.1063/1.333572}}.
\newline\urlprefix\url{https://doi.org/10.1063/1.333572}

\bibitem{doi:10.1063/5.0015977}
D.-N. Nguyen, D.-A. Dao, T.~Miyake, H.-C. Dam, Boron cage effects on
  nd–fe–b crystal structure’s stability, The Journal of Chemical Physics
  153~(11) (2020) 114111.
\newblock \href {http://dx.doi.org/10.1063/5.0015977}
  {\path{doi:10.1063/5.0015977}}.

\bibitem{Okeeffe_coordination_number}
M.~O'Keeffe, A proposed rigorous definition of coordination number, Acta. Cryst
  A35 (1979) 772.

\bibitem{Schmidhuber_2015}
J.~Schmidhuber, \href{http://dx.doi.org/10.1016/j.neunet.2014.09.003}{Deep
  learning in neural networks: An overview}, Neural Networks 61 (2015)
  85–117.
\newblock \href {http://dx.doi.org/10.1016/j.neunet.2014.09.003}
  {\path{doi:10.1016/j.neunet.2014.09.003}}.
\newline\urlprefix\url{http://dx.doi.org/10.1016/j.neunet.2014.09.003}

\bibitem{adagrad}
J.~C. Duchi, E.~Hazan, Y.~Singer, Adaptive subgradient methods for online
  learning and stochastic optimization., J. Mach. Learn. Res. 12 (2011)
  2121--2159.

\bibitem{chollet2015keras}
F.~Chollet, et~al., \href{https://github.com/fchollet/keras}{Keras} (2015).
\newline\urlprefix\url{https://github.com/fchollet/keras}

\bibitem{model_multivariates}
A.~J. Izenman (Ed.), Modern multivariate statistical techniques: Regression,
  classification, and manifold learning, CRC Press, 2009.

\end{thebibliography}
	
\end{document}